\documentclass[twocolumn,superscriptaddress,showpacs,prl,amsmath,amssymb]{revtex4}
\usepackage{graphicx}% Include figure files
\usepackage{bm}% bold math
\begin{document}

\title{Mapping spatial persistent large deviations of 
nonequilibrium surface growth processes onto the temporal 
persistent large deviations of stochastic random walk processes}

\author{ M. Constantin}
\affiliation{
Condensed Matter Theory Center,
Department of Physics, University of Maryland, College Park, 
Maryland 20742-4111
}
\affiliation{Materials Research Science and Engineering Center,
Department of Physics, University of Maryland, College Park, 
Maryland 20742-4111
}
\author{ S. \surname{Das Sarma}}
\affiliation{
Condensed Matter Theory Center,
Department of Physics, University of Maryland, College Park, 
Maryland 20742-4111
}

\begin{abstract}
Spatial persistent large deviations probability of surface growth
processes governed by the Edwards-Wilkinson dynamics, $P_x(x,s)$, 
with $-1 \leq s \leq 1$ is mapped isomorphically onto the temporal
persistent large deviations probability $P_t(t,s)$ associated with the 
stochastic Markovian random walk problem. We show using numerical
simulations that the infinite family of spatial persistent large 
deviations exponents $\theta_x(s)$ characterizing the power law 
decay of $P_x(x,s)$ agrees, as predicted on theoretical grounds by
Majumdar and Bray [Phys. Rev. Lett. {\bf 86}, 3700 (2001)] with the 
numerical measurements of $\theta_t(s)$, the continuous family of 
exponents characterizing the long time power law behavior of
$P_t(t,s)$. We also discuss the simulations of the spatial persistence 
probability corresponding to a discrete model in the Mullins-Herring 
universality class, where our discrete simulations do not agree 
well with the theoretical predictions perhaps because of the
severe finite-size corrections which are known to strongly inhibit the
manifestation of the asymptotic continuum behavior in discrete models
involving large values of the dynamical exponent and the associated
extremely slow convergence to the asymptotic regime.  
\end{abstract}


\maketitle

Non-Markovian Gaussian stochastic processes are very widely 
encountered in a large variety of nonequilibrium physical problems 
\cite{stoch}. Considerable theoretical \cite{FPS_book} and experimental
\cite{1exp,exp_dan1,exp_dan2,exp_px} efforts have recently been
devoted to understanding the first-passage statistics in such 
nonequilibrium systems. Recent work \cite{maj} has revealed that 
the time-dependent history of the non-Markovian processes can 
be described by a nontrivial exponent, $\theta_t$, called the 
persistence exponent, which depends on the dimensionality of 
the problem, $d$, and the precise details of the non-Markovian nature
of the underlying stochastic dynamics characterizing the phenomenon. 
The persistence exponent $\theta_t$ describes the asymptotic 
power law decay of the persistence 
probability [$P_t(t) \propto t^{-\theta_t}$] which measures the 
probability that a stochastic variable has not changed its 
characteristics up to time $t$. As a consequence, $\theta_t$ 
provides useful quantitative predictions concerning the 
temporal evolution characteristics of a given stochastic system. 
Once the persistence probability behavior is found, one can 
immediately calculate the asymptotic behavior of the 
first-passage probability, $F(t)$, 
which represents the distribution of the time when the stochastic 
variable under consideration $first$ reaches a fixed reference value:
$F(t)=-dP_t(t)/dt$. In addition, one can also obtain the mean 
first-passage time which provides the representative time scale 
characterizing the stability of the dynamical process. Such a time
scale might be of interest for the study of the evolution of 
fluctuating interfaces or for undestanding the behavior of a
collection of stochastic spin variables. To be specific, we mention 
that $P_t(t)$ and $\theta_t$ are the $temporal$ persistence 
probability and exponent, respectively, since we also discuss 
$spatial$ persistence $P_x(x)$ and the corresponding exponent 
$\theta_x$.

Of particular interest in the field of surface growth phenomena is 
the role played by the dynamics of interfaces which are governed by 
thermal fluctuations. An illustrative category of such interfaces is 
described by linear Langevin equations of the type 
\begin{equation}
\label{EW}
\partial h(x,t)/ \partial t=-(-\nabla^{2})^{z/2}h(x,t)+\xi(x,t), 
\end{equation} 
\noindent where $h(x,t)$ is the step height fluctuation corresponding
to the lateral step position $x$, at time $t$, $\xi(x,t)$ is a white 
uncorrelated Gaussian noise, and $z$ is the dynamical exponent. 
It turns out that fluctuating interfaces are 
of crucial importance at very small scales (i.e. nanoscales) involved 
in the fabrication of current electronic devices. In addition to the 
traditional way of analyzing various aspects of growth processes based 
on the dynamical scaling behavior of the interface width and temporal 
and spatial correlation functions \cite{barabasi,ellen}, it has 
been shown that persistence properties provide an additional tool of 
investigation for understanding the long time evolution of surface 
growth phenomena, due to the ability of the nontrivial persistence 
exponents to identify the underlying universality class of the 
dynamical process \cite{krug} and the presence of the nonlinearities 
associated with the dynamical evolution \cite{magda_Pt}. However, 
much broader and more general information can be extracted from the 
natural generalization of the persistence through the probability 
of persistent large deviations \cite{DG}, $P_t(t,s)$, where 
$-1 \leq s \leq 1$. A closely related concept, the sign-time
distribution, has been introduced in Ref.~\cite{zoltan}. 
%+++++++++++++++++++++ new:
$P_t(t,s)$ measures the probability that the average sign, 
$S_{\text{av}}(t)=(1/t) \int_{0}^{t} \hbox{sgn} 
[{h(x,t_0+t^{\prime})-h(x,t_0)}]~dt^{\prime}$, remains always 
above a particular value $s$ up to time $t$ measured from the 
initial time $t_0$. 
%+++++++++++++++++++++
It turns out that $P_t(t,s)$ provides an 
infinite family of temporal persistence exponents, $\theta_t(s)$, 
associated with the power law decay of $P_t(t,s)$ observed to exist at 
large time scales for systems belonging to different universality 
classes \cite{magda_Pts} relevant for surface growth. In the limit 
$s \rightarrow 1$, the exponent of persistent large deviations 
reaches the value of the nontrivial persistence exponent, i.e., 
$\theta_t(s=1)=\theta_t$. We note that the concept of persistent large 
deviations naturally generalizes the concept of persistence exponent 
from a single discrete exponent $\theta_t$ characterizing the 
universality class to a more general and deeper concept of a 
continuous function, $\theta_t(s)$, of the persistent exponents 
characterizing the stochastic dynamics.

The dynamics of the spatially extended systems with fluctuations governed 
by stochastic differential equations can be further elucidated by looking, 
in addition to the statistical tools mentioned above, at the $spatial$ 
analog of the temporal persistence, i.e., the spatial persistence 
probability \cite{maj_px,magda_Px}, $P_x(x)$, and its associated exponents. 
$P_x(x)$ represents the probability that the height stochastic variable, 
measured at a fixed time $t$, does not reach its initial value $h(x_0,t)$ 
up to a longer distance $x$ measured from the initial position
$x_0$. Theoretical \cite{maj_px} and numerical studies 
\cite{magda_Px} indicate that the power law decay of $P_x(x)$ is
described by two different exponents, the steady-state
($\theta_{SS}$) and the finite initial conditions ($\theta_{FIC}$)
spatial persistence exponents depending on the selection rules 
applied to $x_0$: (i) $\theta_{SS}$ is obtained if $x_0$ is 
sampled from the entire set of the steady-state configurational 
sites, and (ii) $\theta_{FIC}$ is obtained if $x_0$ is sampled 
from a subset of the steady-state sites characterized by 
$finite$ height and height derivatives. The aim of this paper 
is to establish numerically the concept of $spatial$ persistent 
large deviations probability, $P_x(x,s)$ with $-1 \le s \le 1$, 
as a natural generalization of the spatial persistence 
probability concept. We also show that $P_x(x,s)$ measured for growth 
processes in the well-studied Edwards-Wilkinson \cite{EWuniv} 
universality class [described by Eq.~(\ref{EW}) with $z=2$] can be 
mapped isomorphically onto $P_t(t,s)$ of the simple 
random walk stochastic problem. This mapping possibility is 
inspired by the work of Majumdar and Bray \cite{maj_px}, who 
have shown in a recent Letter that the $spatial$ persistence 
probability characteristics of growth processes involving the 
interfacial height stochastic variable $h(x,t)$ with the 
dynamics described by Eq.~(\ref{EW}) can be mapped onto the 
$temporal$ persistence characteristics of the 
``random walk'' processes of the type 
$\hbox{d}^n x/\hbox{d}t^n= \eta(t)$, where $n=(z-d+1)/2$ 
and $\eta(t)$ is a white noise as well. The purpose of the current
paper is to show that this exact mapping, as expected, works for the
generalized (large deviations) persistence probability and the
corresponding continuous family of persistence exponents as well, 
and to numerically calculate $\theta_x(s)$ for the important
class of processes controlled by the Edwards-Wilkinson equation.

We consider the average sign of the interfacial height stochastic 
variable measured at a fixed time $t$ with respect to the original 
value corresponding to the initial position $x_0$,
\begin{equation}
\label{avr_sign}
S_{\hbox{av}}(x) = \frac{1}{x} \int_{0}^{x} \hbox{sgn} 
[{h(x_0+x^{\prime},t)-h(x_0,t)}]~dx^{\prime} .
\end{equation}
The spatial persistent large deviations probability is defined, 
in analogy with its temporal correspondent, as the probability 
that the average sign $S_{\text{av}}$ remains persistently 
above a particular value $s$, with $-1 \le s \le 1$, up to a longer 
distance $x$ measured from the initial position $x_0$, 
\begin{equation}
P_x(x,s) \equiv \hbox{Prob}~\lbrace ~S_{\text{av}}(x^{\prime}) \geq s,
~\forall x^{\prime} \leq x~ \rbrace.
\end{equation}
\noindent We provide numerical evidence showing that $P_x(x,s)$ 
has a power law behavior for $x<L$, where $L$ is the typical 
length scale in the numerical simulations (i.e. system size), 
independent of the choice of the average sign parameter $s$,
\begin{equation}
\label{Pxs}
P_x(x,s) \propto x^{-\theta_x(s)}, 
\end{equation}
\noindent where the spatial persistent large deviations exponent 
$\theta_x(s)$ depends continuously on the parameter $s$ that appears 
in the definition of the probability. The importance of $P_x(x,s)$ 
lies in the fact that it provides an $infinite$ family of persistence 
exponents, instead of only one exponent as in the case of 
$P_x(x)$. Obviously, $P_x(x,s=1)$ and its associated exponent 
$\theta_x(s=1)$ are precisely the spatial persistence probability 
and the nontrivial persistence exponent ($\theta_{SS}$ or
$\theta_{FIC}$, depending on the sampling procedure applied to $x_0$), 
respectively. The opposite limit $s \to -1$ is trivial 
in the sense that $P_x(x,s=-1)=1$ independent of $x$ and as a 
consequence $\theta_x(s=-1)=0$. 

The $s$ dependence of the temporal persistent large devations
exponents is known exactly for the simple random walk case, which is
one of the few analytically solved persistence problems \cite{th_Pts},
\begin{equation}
\label{th_t_s}
\theta_t(s)=\frac{2\theta_t(1)}{\pi} \arctan{\sqrt{\frac{1+s}{1-s}}}.
\end{equation}
\noindent The mapping \cite{maj_px} between the $temporal$ properties 
of the random walker (RW) problem and the $spatial$ properties of the 
Edwards-Wilkinson (EW) fluctuating interfaces implies that the 
expression of Eq.~(\ref{th_t_s}) also applies to the distribution 
of the $spatial$ persistent large deviations exponent as a function 
of $s$. This conjecture is verified numerically in this study.

In this paper we have carried out the first application of the spatial
persistent large deviations concept to the case of (1+1)-dimensional
fluctuating interfaces characterized by the EW dynamical
equation. Using the configuration of the interface corresponding to 
a fixed time of the order of the time required by the interface width 
to saturate (i.e., $t \sim L^z$), we have computed $P_x(x,s)$ as the 
fraction of lattice sites $x_j$ (with $j=1,2,...,L-1$) which
maintained their stochastic variable $S_{\hbox{av}}$ persistently 
above a fixed $s$ value, up to a distance $x_j+x$. The initial
measurement points, $x_j$, are sampled from the entire set of the
steady-state interfacial profile. The numerical integration of the 
stochastic equation is performed using the simple 
forward-time centered-space representation \cite{num_rec}. 
We have used numerical systems of size $L \sim 1000$ and we have 
averaged the results over many ($\sim 1000$) independent runs to
obtain convergent statistics. 

\begin{figure}
\includegraphics[height=7.5cm,width=7.7cm]{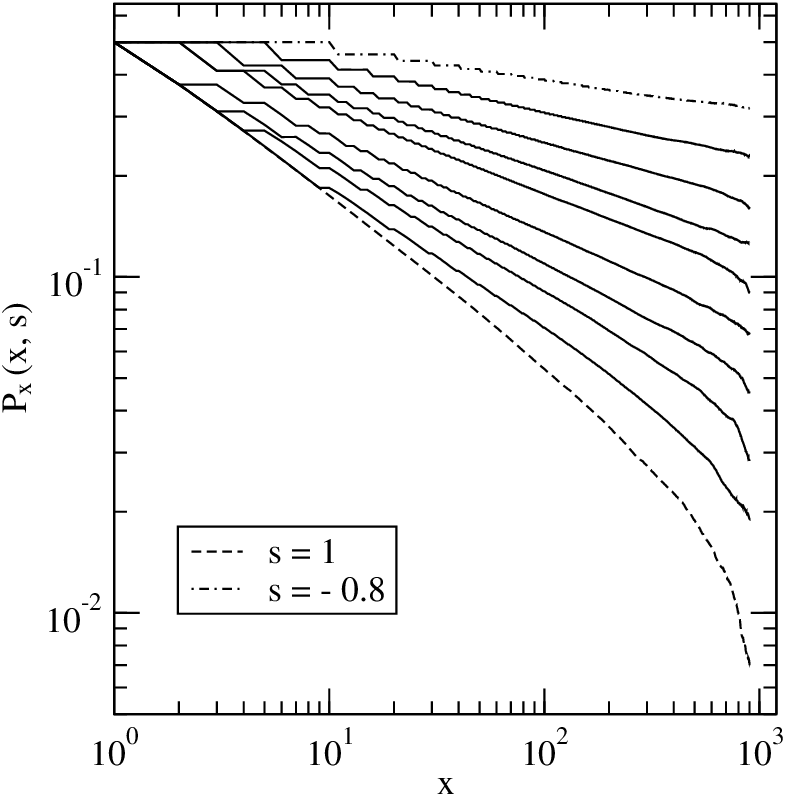} %includes EW_pxs.eps
\caption{\label{fig1} Log-log plot of $P(x,s)$ versus $x$ for the EW 
equation based on the direct numerical integration of Eq.~(\ref{EW}) 
with $z=2$, using a system of size $L=1000$. The average sign parameter takes 
ten different values decreasing from $s=1$ (bottom curve) to $s=-0.8$ 
(top curve) with an average sign difference $\Delta s=0.2$. All 
spatial persistent large deviations probabilities show power law 
decay vs distance for $x<L/2$. The finite-size effects are
responsible for the deviations of the probabilities from the power law
trend at large values of $x$.}
\end{figure}

In Fig.~\ref{fig1} we show the results for $P_x(x,s)$ as a function of
$x$ for (1+1)-dimensional EW interfaces simulated numerically. 
We display ten log-log spatial persistent large deviations curves versus
the distance $x$ for ten values of the average sign parameter $s$
(i.e., $s=+1,~+0.8,~\ldots,~-0.8$). We observe that 
$P_x(x,s) \sim x^{-\theta_x(s)}$ for $x<L/2$, while for larger values of
$x$ and $s \geq 0$ there is a downward deviation of the probability 
from the power law behavior due to finite-size limitations. 
Except for the curve corresponding to $s=1$, which gives the usual 
spatial persistence exponent $\theta_{SS} \simeq 0.50$, in agreement 
with Refs.~\cite{maj_px,magda_Px}, all the other curves with 
$s<1$ provide new information concerning spatial behavior of the 
interface fluctuations.

\begin{figure}
\includegraphics[height=7.5cm,width=7.5cm]{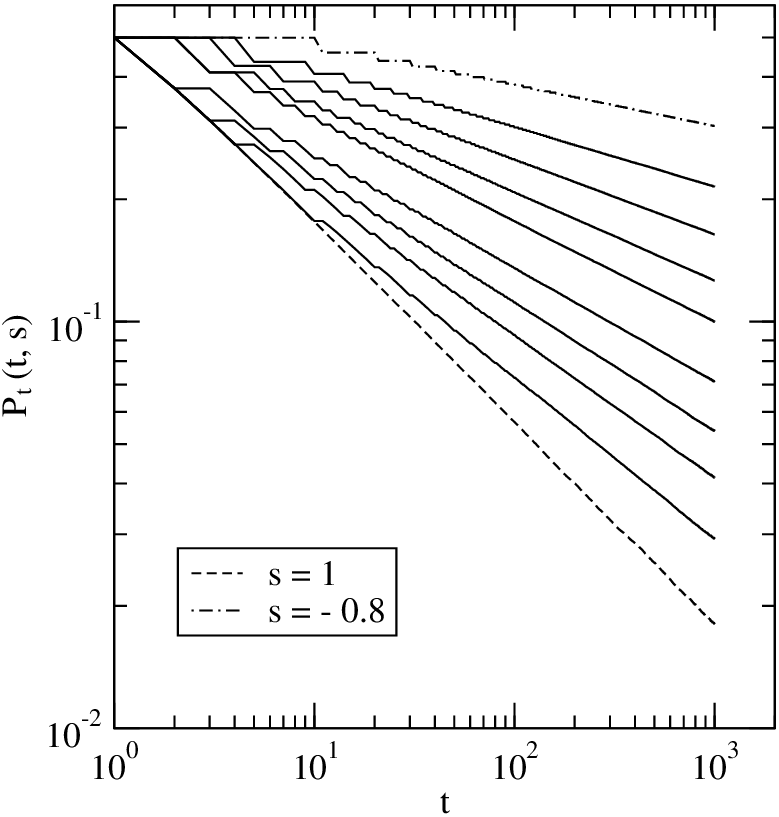} %includes RW.eps
\caption{\label{fig2} Log-log plot of simulated $P_t(t,s)$ vs $t$ 
for the RW problem. The system size is $L=500$ and the 
average sign parameter takes ten different values decreasing from 
$s=1$ (bottom curve) to $s=-0.8$ (top curve) with $\Delta s=0.2$ 
between successive probability curves. All temporal persistent 
large deviations probabilities show power law behavior vs time.}
\end{figure}

\begin{figure}
\includegraphics[height=7cm,width=8.2cm]{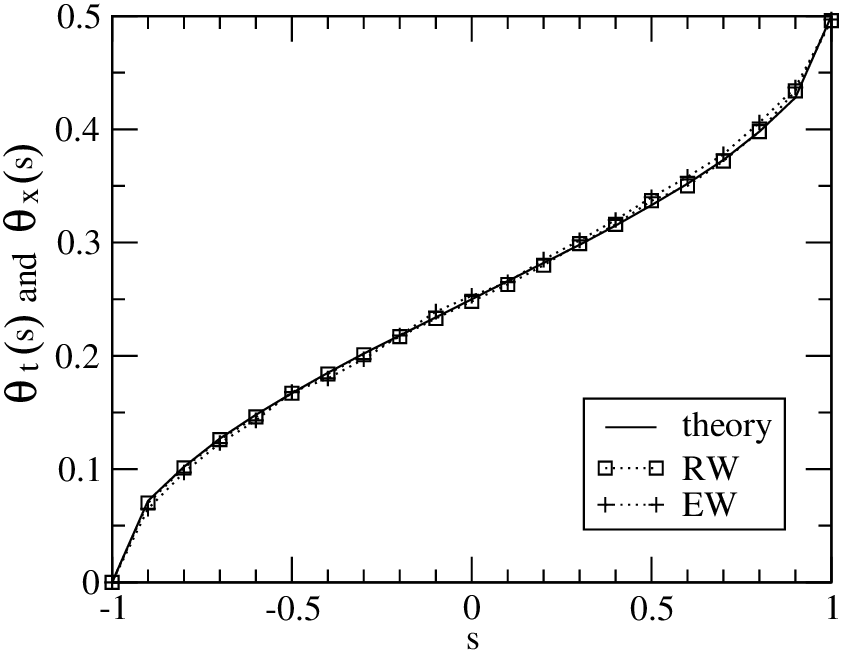}
\caption{\label{fig3} $\theta_t(s)$ and $\theta_x(s)$ vs $s$ as 
extracted from the power-law decay of $P(t,s)$ (for the RW problem) 
and $P(x,s)$ (for the EW fluctuating interfaces), respectively. 
The increment of the average sign parameter is $\Delta s=0.1$. 
The continuous curve represents the theoretical prediction of 
Eq.~(\ref{th_t_s}).}
\vspace{1.5cm}
\end{figure}

The temporal persistent large deviations probability of the random
walk model is shown in Fig.~\ref{fig2}. We have used similar $s$
values, as in the case of $P_x(x,s)$ described above. $P_t(t,s)$ shows
a clear power law behavior versus $t$. We find that $P_t(t,s=1)$ is
characterized by an exponent of $~0.50$, in agreement with the
theoretical value $\theta_t=1/2$. Individual temporal persistent large
deviations exponents $\theta_t(s)$ are extracted from the linear 
regions of the log-log plots of $P_t(t,s)$ versus $t$ and they 
are compared in Fig.~\ref{fig3} to the corresponding spatial set 
of exponents $\theta_x(s)$ for the EW interfaces. 

The level of agreement between $P_x(x,s)$ corresponding to the EW
dynamical equation and $P_t(t,s)$ of the RW case can be seen 
in Fig.~\ref{fig3}. To generate this figure we have used an increment
of the average sign parameter ($s$) of $0.1$. We observe that the two 
sets of exponents, $\theta_t(s)$ and $\theta_x(s)$, overlap very 
well within the errors of our simulations, showing that the mapping
procedure involved in this study is perfectly applicable. Both cases 
are in agreement with the theoretical prediction of
Eq.~(\ref{th_t_s}). We have also simulated a discrete stochastic
growth model, the so-called Family model, which is theoretically
known to exactly belong to the EW universality class
\cite{magda_Pt}. The Family model results (not shown here) for $P_x(x,s)$ and
$\theta_x(s)$ are very similar to those shown in Fig.~\ref{fig1} since
they have identical stochastic dynamics. 

\begin{figure}
\includegraphics[height=6cm,width=8.2cm]{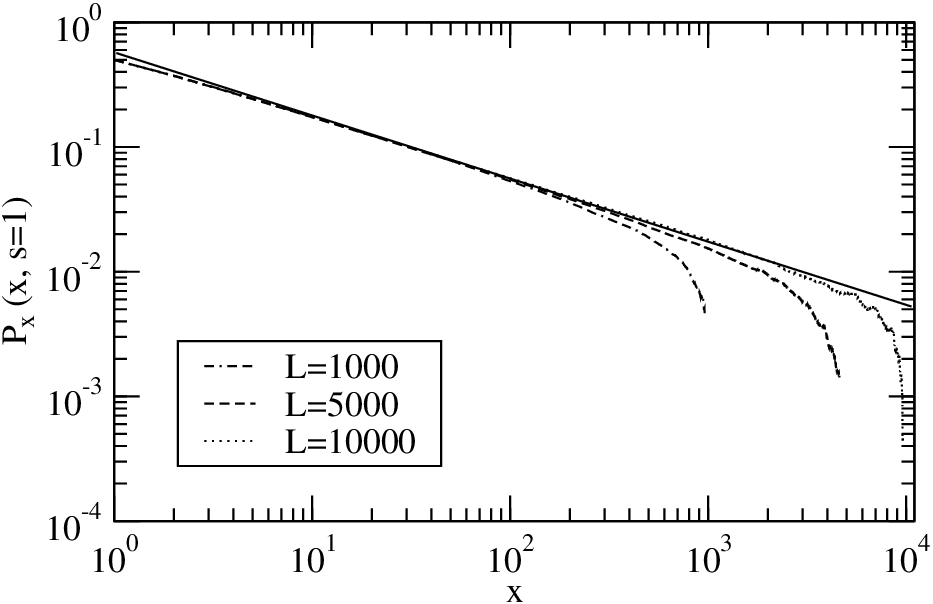} %includes diff_L.eps
\caption{\label{fig4} Log-log plot of $P_x(x,s)$ for $s=1$ 
corresponding to the EW equation based on the direct numerical 
integration of Eq.~(\ref{EW}) with $z=2$, using three system sizes, 
as shown in the legend. The straight line represents the fit for 
$L=10^4$ simulation, providing an exponent of $1/2$.}
\end{figure}

%++++++++++++++++++++++++++new
Despite the downward deviation of the probability $P_x(x,s)$ from the 
power law behavior due to finite-size limitations, we have checked
that larger system sizes would provide a wider range of distances over
which the spatial persistence large deviations exponent can be
extracted with a better precision. This can be seen in Fig.~\ref{fig4}. 

%+++++++++++++++++++++++++++++++

%
Another case of interest for epitaxial surface dynamics is growth 
under surface diffusion minimizing the local curvature, which belongs 
asymptotically to the Mullins-Herring (MH) \cite{MH} universality class 
[i.e. Eq.~(\ref{EW}) with $z=4$]. The exact mapping prediction by 
Majumdar and Bray \cite{maj_px} suggests that the spatial persistence 
properties of the $continuum$ version of the growth models 
belonging to this universality class could be mapped onto 
the temporal persistence characteristics of the random 
acceleration problem described by the stochastic random equation 
$\hbox{d}^2 x/\hbox{d}t^2= \eta(t)$ with an analytically known
exponent of $\theta_t=1/4$ \cite{random_acc}. One expects to obtain 
$\theta_{SS}=0$ and $\theta_{FIC}=1/4$ \cite{maj_px} when measuring the
steady-state and finite initial conditions regimes of $P_x(x)$, 
respectively, for the Mullins-Herring surface growth dynamics. 
%added++++
An example of this case is the (1+1)-dimensional model 
introduced by Kim and Das Sarma \cite{KD}. This discrete
solid-on-solid atomistic model, the so-called larger curvature 
(LC) model \cite{KD} is known to belong asymptotically to the MH 
universality class.
%%%ok
As a consequence, we focus on the measurement of
$P_{FIC}(x)$ for the discrete LC model \cite{KD}, 
since $P_{SS}(x)$ is trivially described by a 
null exponent. The definition of $P_{FIC}(x)$ involves the selection
of the subset of sites characterized by $finite$ height and height 
derivatives. One possibility would be to sample over the subset of sites
placed on the average level. However, it turns out that a system with
$L=200$, which is the typical system size in our simulations, 
usually has only a couple of discrete positions on the
average level. For this reason, we have sampled over 
all the lattice sites $x_{j}$ with the height variable (measured with respect
to the average level) within a band of values characterized by a 
width $w$ [i.e., $-w/2 \leq h(x_j) \leq w/2$], where $w$ is taken to be
smaller than the maximum magnitude of interface fluctuations. 
This selection ensured the possibility of sampling over a reasonable 
number of lattice sites presumably sufficient for good statistical 
results of $P_{FIC}(x)$.

\begin{figure}
\includegraphics[height=6.3cm,width=8.2cm]{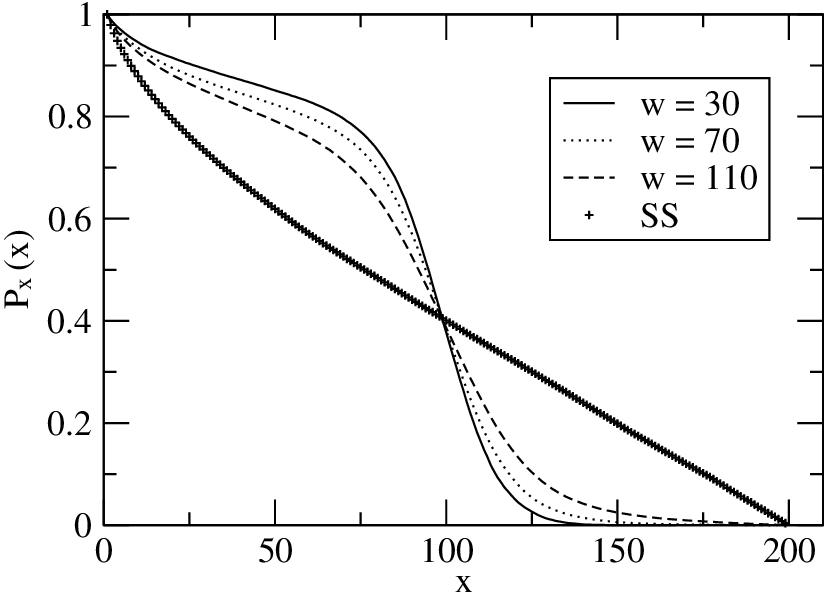} %includes LC.eps
\caption{\label{fig5} Numerical results of $P_{FIC}(x)$ and 
$P_{SS}(x)$ for the LC discrete model with system size $L=200$. 
The measurements are performed from steady-state configurations. 
$P_{FIC}(x)$ probabilities have been obtained by using three 
different band widths, as shown in the legend. $P_{FIC}(x)$ does not 
display a power-law behavior as a function of $x$ over the entire 
range of system size. }
\end{figure}

In Fig.~\ref{fig5} we show the $x$ dependence of $P_{FIC}(x)$ 
corresponding to the LC discrete model for three values
of $w$: 30, 70 and 110, respectively. The steady-state probability is
shown for comparison. We note that $P_{FIC}(x)$ does not display the
expected power-law behavior as a function of $x$. As the bandwidth
$w$ increases, more and more lattice sites are included in the 
sampling subset, and $P_{FIC}(x)$ tends to reach the behavior
displayed by $P_{SS}(x)$. In addition, we observe that when using a
numerical system with $L=200$, $P_{SS}(x)$ has a rather linear
dependence on $x$, for $50<x<200$. The impossibility to recover the
theoretically predicted behavior of $P_{FIC}(x)$ may be due to 
the reduced system size used in our simulations. This limitation 
is imposed by the requirement of measuring the probability
$P_{FIC}(x)$ using an ensemble of steady-state configurations that can
be achieved only by using an extensive computational time $\sim L^4$.
%++++++++++++++++++++++++
We note that reducing or increasing the system size by a factor of 2
did not produce any qualitative change in the overall behavior of
$P_{FIC}(x)$ or $P_{SS}(x)$. In addition, we have checked that the 
direct numerical integration of Eq.~(\ref{EW}) with $z=4$ provides 
results consistent with the discrete LC model. Also, it turns out 
that similar probability curves are obtained for solid-on-solid models 
belonging asymptotically to the molecular beam epitaxy universality 
class (such as the (1+1)-dimensional DT model \cite{DT}). 
%++++++++++++++++++++++++
We believe that our problem with the spatial persistence
$P_x(x)$ for the LC model belonging to the MH universality 
class \cite{KD} arises most likely from the severe finite-size 
problems in simulating systems with large values ($z=4$) 
of the dynamical exponent. Large dynamical exponent implies very slow
lateral correlations, which considerably complicates studying
steady-state behavior in the MH universality problem.
%add also this+++++++++++++
In fact, this issue is very well known in traditional studies 
of dynamical scaling involving surface phenomena characterized by a
large value of the dynamical exponent \cite{dyn_sc}. A large variety
of stochastic discrete models show long-time transients and they cross
over very slowly to their corresponding asymptotic behavior. 
Only extensive simulations of stochastic discrete models in 
the MH universality class can provide the asymptotic dynamical 
scaling associated with the continuous limit of the Eq.~(\ref{EW})
with $z=4$.     
%done++++++++++++++++++++++
This forbids us from pursuing further measurements of $P_x(x,s)$ for 
the MH universality class and checking the validity 
of the mapping procedure, which remains an interesting open
problem.

To conclude, we have shown numerically that the spatial persistent
large deviations probability represents a possible generalization of
the spatial persistence probability, providing a useful family of
spatial exponents for the surface growth phenomena. We have mapped 
these exponents into the family of temporal persistent large
deviations exponents obtained from the evolution of a simple 
stochastic ``random walk'' process. We have established the validity 
of this generalization for the case of fluctuating interfaces 
described by the Edwards-Wilkinson evolution equation. 
However, the similar problem involving the Mullins-Herring 
universality class remains open since the corresponding discrete LC
model \cite{KD} simulation shows severe finite-size problems.

%======================================================================
This work is partially supported by NSF-DMR-MRSEC and U.S. ONR.
%======================================================================

%======================================================================


\begin{thebibliography}{99}
\bibitem{stoch}C. W. Gardiner, {\it Handbook of Stochastic Methods 
for Physics, Chemistry and the Natural Sciences} (Springer-Verlag, 
Berlin, 1983).
\bibitem{FPS_book} J. Grasman, {\it Asymptotic Methods for the 
Fokker-Planck Equation and the Exit Problem in Applications} 
(Springer-Verlag, Berlin, 1999). 
\bibitem{1exp} M. Marcos-Martin, D. Beysens, J. P. Bouchaud,
C. Godreche, and I. Yekutieli, 
Physica A {\bf 214}, 396 (1995); 
W. Y. Tam, R. Zeitak, K. Y. Szeto, and J. Stavans, 
Phys. Rev. Lett. {\bf 78}, 1588 (1997); 
B. Yurke, A. N. Pargellis, S. N. Majumdar, and C. Sire, 
Phys. Rev. E {\bf 56}, R40 (1997); 
G. P. Wong, R. W. Mair, R. L. Walsworth, and D. G. Cory, 
Phys. Rev. Lett. {\bf 86}, 4156 (2001). 
\bibitem{exp_dan1} D. B. Dougherty, I. Lyubinetsky, E. D. Williams,
M. Constantin, C. Dasgupta, and S. Das Sarma, 
Phys. Rev. Lett {\bf 89}, 136102 (2002).
\bibitem{exp_dan2} D. B. Dougherty, O. Bondarchuk, M. Degawa, 
and E. D. Williams, Surf. Sci. {\bf 527}, L213 (2003).
\bibitem{exp_px} J. Merikoski, J. Maunuksela, M. Myllys, J. Timonen, 
and M. J. Alava, Phys. Rev. Lett. {\bf 90}, 024501 (2003).
\bibitem{maj} For a review, see S. N. Majumdar, Curr. Sci. {\bf 77},
370 (1999); J. Krug, e-print cond-mat/0403267.
\bibitem{barabasi} A. -L. Barabasi and H. E. Stanley, {\it Fractal Concepts
in Surface Growth } (Cambridge University Press, New York, 1995).
\bibitem{ellen} H. Jeong and E. D. Williams, Surf. Sci. Rep. 
{\bf 34}, 171 (1999).
\bibitem{krug} J. Krug, H. Kallabis, S. N. Majumdar, S. J. Cornell,
A. J. Bray, and C. Sire, Phys. Rev. E {\bf 56}, 2702 (1997).
\bibitem{magda_Pt} M. Constantin, C. Dasgupta, P. Punyindu 
Chatraphorn, S. N. Majumdar, and S. Das Sarma, 
Phys. Rev. E {\bf 69}, 061608 (2004).
\bibitem{DG} I. Dornic and C. Godreche, J. Phys. A {\bf 31}, 5413 (1998).
\bibitem{zoltan} T. J. Newman and Z. Toroczkai, Phys. Rev. E {\bf 58}, R2685 (1998);
Z. Toroczkai, T. J. Newman, and S. Das Sarma, Phys. Rev. E {\bf 60}, R1115 (1998).
\bibitem{magda_Pts} M. Constantin, S. Das Sarma, C. Dasgupta,
O. Bondarchuk, D. B. Dougherty and E. D. Williams,
Phys. Rev. Lett. {\bf 91}, 086103 (2003).
\bibitem{maj_px} S. N. Majumdar and A. J. Bray, Phys. Rev. Lett. {\bf 86},
3700 (2001).
\bibitem{magda_Px} M. Constantin, S. Das Sarma, and C. Dasgupta, 
Phys. Rev. E {\bf 69}, 051603 (2004). 
\bibitem{EWuniv} S. F. Edwards and D. R. Wilkinson, Proc. R. Soc. London,
Ser. A {\bf 381}, 17 (1982).
\bibitem{th_Pts} A. Baldassarri, J. P. Bouchaud, I. Dornic, and C. Godreche, 
Phys. Rev. E {\bf 59}, R20 (1999).
\bibitem{num_rec} W. H. Press {\it et al.}, {\em  Numerical Recipes}
`(Cambridge University Press, Cambridge, 1989).
\bibitem{random_acc} T. W. Burkhardt,  J. Phys. A {\bf 26}, L1157 (1993);
Y.G. Sinai, Theor. Math. Phys. {\bf 90}, 219 (1992).
\bibitem{MH} W. W. Mullins, J. Appl. Phys. {\bf 28}, 333 (1957); C. Herring,
J. Appl. Phys. {\bf 21}, 301 (1950).
\bibitem{KD} J. M. Kim and S. Das Sarma, Phys. Rev. Lett. {\bf 72}, 2903 (1994); 
J. Krug, Phys. Rev. Lett. {\bf 72}, 2907 (1994).
\bibitem{DT} S. Das Sarma and P. Tamborenea, Phys. Rev. Lett. {\bf
  66}, 325 (1991).
\bibitem{dyn_sc} S. Das Sarma, C. J. Lanczycki, R. Kotlyar, and
S. V. Ghaisas, Phys. Rev. E {\bf 53}, 359 (1996); P. Punyindu and
S. Das Sarma, Phys. Rev. E {\bf 57}, R4863 (1998).
\end{thebibliography}
\end{document}